\documentclass[12pt]{article}
\usepackage{amsmath,amsfonts,amssymb}

\begin{document}
\thispagestyle{empty}

\def\theequation{\arabic{section}.\arabic{equation}}
\def\a{\alpha}
\def\b{\beta}
\def\g{\gamma}
\def\d{\delta}
\def\dd{\rm d}
\def\e{\epsilon}
\def\ve{\varepsilon}
\def\z{\zeta}
\def\B{\mbox{\bf B}}\def\cp{\mathbb {CP}^3}

\newcommand{\h}{\hspace{0.5cm}}

\begin{titlepage}
\vspace*{1.cm}
\renewcommand{\thefootnote}{\fnsymbol{footnote}}
\begin{center}
{\Large \bf String solutions in $AdS_3\times S^3\times S^3\times
S^1$ with $B$-field}
\end{center}
\vskip 1.2cm \centerline{\bf Plamen Bozhilov} \vskip 0.6cm
\centerline{\sl Institute for Nuclear Research and Nuclear Energy}
\centerline{\sl Bulgarian Academy of Sciences} \centerline{\sl
1784 Sofia, Bulgaria}

\centerline{\tt bozhilov@inrne.bas.bg, bozhilov.p@gmail.com}

\vskip 20mm

\begin{center}
{\bf Abstract}
\end{center}

\h We consider strings living in $AdS_3\times S^3\times S^3\times
S^1$ with nonzero $B$-field. By using specific ansatz for the
string embedding, we obtain a class of solutions corresponding to
strings moving in the whole ten dimensional space-time. For the
 $AdS_3$ subspace, these solutions are given in terms of
 incomplete elliptic integrals. For the two three-spheres, they
 are expressed in terms of Lauricella hypergeometric functions of
 many variables. The conserved charges, i.e. the string energy,
 spin and angular momenta, are also found.

\end{titlepage}

\newpage
\baselineskip 18pt

%%%%%%%%%%%%%%%%%%%%%%%%%%%%%%%%%%%%%%%%%%%%%%%%%
\def\nn{\nonumber}
%%%%%%%%%%%%%%%%%%%%%%%%%%%%%%%%%%%%%%%%%%%%%%%%%
%%%%%%%%%%%%%%%%%%%%%%%%%%%%%%%%%%%%%%%%%%%%%%%%%%%%%
\def\tr{{\rm tr}\,}
\def\p{\partial}
\newcommand{\non}{\nonumber}
\newcommand{\bea}{\begin{eqnarray}}
\newcommand{\eea}{\end{eqnarray}}
\newcommand{\bde}{{\bf e}}
\renewcommand{\thefootnote}{\fnsymbol{footnote}}
\newcommand{\be}{\begin{eqnarray}}
\newcommand{\ee}{\end{eqnarray}}
%\newcommand{\h}{\hspace{0.5cm}}
%%%%%%%%%%%%%%%%%%%%%%%%%%%%%%%%%%%%%%%%%%%%%%%%%%%%

\vskip 0cm

\renewcommand{\thefootnote}{\arabic{footnote}}
\setcounter{footnote}{0}

\setcounter{equation}{0}
%%%%%%%%%%%%%5%%%%%%%%%%%%%%%%%%%%%%%%%%%%%%%%%%%%%%
\section{Introduction}
%%%%%%%%%%%%%5%%%%%%%%%%%%%%%%%%%%%%%%%%%%%%%%%%%%%%

A very important development in the field of string theory has
been achieved for the case of AdS/CFT duality \cite{AdS/CFT}
between strings and conformal field theories in various
dimensions. The most developed case is the correspondence between
strings living in $AdS_5\times S^5$ and $\mathcal{N}=4$ SYM in
four dimensions. Another example is the duality between strings on
$AdS_4\times CP^3$ background and $\mathcal{N}=6$ super
Chern-Simons-matter theory in three space-time dimensions. The
main achievements in the above examples are due to the discovery
of integrable structures on both sides of the correspondence. Many
other cases have been considered also \cite{RO}.

An interesting area of research is the $AdS_3/CFT_2$ duality
\cite{BSZ0912}-\cite{BPS1508}, related to $AdS_3\times S^3\times
T^4$ and $AdS_3\times S^3\times S^3\times S^1$ string theory
backgrounds where nontrivial two-form $B$ field appears. For a
review, see e.g \cite{Sfondrini}.

The classical string solutions and their semiclassical limits,
corresponding to large conserved charges \cite{GKP0204051}, play
important role in checking and understanding the AdS/CFT
correspondence. Here we obtain a class of solutions corresponding
to strings moving in the whole ten dimensional space-time
$AdS_3\times S^3\times S^3\times S^1$ with nonzero $B$-field.

The paper is organized as follows. In Sec.2 we describe the
background. In Sec.3 we present our general approach to string
dynamics in curved backgrounds with nonzero B-field. In Sec.4 we
apply it to strings moving in $AdS_3\times S^3\times S^3\times
S^1$ with $B$-field. In Sec.5 we obtain the conserved charges for
the case under consideration. Sec.6 is devoted to our concluding
remarks.

\setcounter{equation}{0}
%%%%%%%%%%%%%5%%%%%%%%%%%%%%%%%%%%%%%%%%%%%%%%%%%%%%
\section{The background}
%%%%%%%%%%%%%5%%%%%%%%%%%%%%%%%%%%%%%%%%%%%%%%%%%%%%

The metric of $AdS_3\times S^3\times S^3\times S^1$ is
\bea\label{m} ds^2=ds^2_{AdS_3}+ds^2_{S^3_+}
+ds^2_{S^3_-}+dw^2,\eea where $w$ is the coordinate along $S^1$.
As found in \cite{GMT98}, the radii of $AdS_3$ and of the two
three-spheres satisfy the relation \bea\label{rr}
\frac{1}{R^2_{AdS_3}}=\frac{1}{R^2_+}+\frac{1}{R^2_-}.\eea If we
normalize the $AdS_3$ radius to one, (\ref{rr}) is solved by \bea
\frac{1}{R^2_+}=\cos^2 \varphi,\h \frac{1}{R^2_-}=\sin^2 \varphi
.\eea

According to \cite{BSSS15}, the metric on $AdS_3$ and the two
three-spheres can be written as \bea\label{bA} &&ds^2_{AdS_3} =
-\left(\frac{1+\frac{z_1^2+z_2^2}{4}}{1-\frac{z_1^2+z_2^2}{4}}\right)^2
dt^2+\left(\frac{1}{1-\frac{z_1^2+z_2^2}{4}}\right)^2
(dz_1^2+dz_3^2),
\\ \label{p} && ds^2_{S^3_+} =
\left(\frac{1-\cos^2 \varphi\frac{y_3^2+y_4^2}{4}}{1+\cos^2
\varphi\frac{y_3^2+y_4^2}{4}}\right)^2 d\phi_5^2 +
\left(\frac{1}{1+\cos^2 \varphi\frac{y_3^2+y_4^2}{4}}\right)^2
(dy_3^2+dy_4^2),
\\ \label{mm} &&  ds^2_{S^3_-} =
\left(\frac{1-\sin^2 \varphi\frac{x_6^2+x_7^2}{4}}{1+\sin^2
\varphi\frac{x_6^2+x_7^2}{4}}\right)^2 d\phi_8^2 +
\left(\frac{1}{1+\sin^2 \varphi\frac{x_6^2+x_7^2}{4}}\right)^2
(dx_6^2+dx_7^2) .\eea

The $B$-field in these coordinates is given by \cite{BSSS15}
\bea\label{Bf} &&B=
\frac{q}{\left(1-\frac{z_1^2+z_2^2}{4}\right)^2} (z_1 dz_2-z_2
dz_1)\wedge dt \\ \nn &&+\frac{q\cos\varphi}{\left(1+\cos\varphi \
\frac{y_3^2+y_4^2}{4}\right)^2} (y_3 dy_4-y_4 dy_3)\wedge d\phi_5
\\ \nn &&+ \frac{q \sin\varphi}{\left(1+\sin\varphi \ \frac{x_6^2+x_7^2}{4}\right)^2}
(x_6 dx_7-x_7 dx_6)\wedge d\phi_8,\eea where the parameter $q$ is
related to the quantized coefficient $k$ of the Wess-Zumino term
by \cite{BSSS15} \bea\label{k} k=q\sqrt{\lambda}.\eea

For our purposes here, we introduce new background coordinates:
\bea\label{Z} &&z_1=2 \tanh \frac{\rho}{2} \cos\phi,\h z_2=2 \tanh
\frac{\rho}{2}\sin \phi,
\\ \label{Y} &&\phi_5= R_+ \phi_{2+} \\ \nn &&
y_3=R_+ w_1= 2R_+ \tan\frac{\theta_+}{2}\cos \phi_{1+}\\ \nn
&&y_4=R_+ w_2 = 2R_+ \tan\frac{\theta_+}{2} \sin \phi_{1+},
\\ \label{X} &&\phi_8= R_- \phi_{2-}
\\ \nn &&
x_6=R_- v_1= 2R_- \tan\frac{\theta_-}{2}\cos \phi_{1-}
\\ \nn
&&x_7=R_- v_2 = 2R_- \tan\frac{\theta_-}{2} \sin \phi_{1-} . \eea
As a consequence, the resulting description of the background
becomes: \bea\label{bAt} &&ds^2_{AdS_3} =
 -\cosh^2\rho \ dt^2+d\rho^2+\sinh^2\rho \ d\phi^2, \eea
or ($\sinh^2\rho=r^2$) \bea\label{bAtr} &&ds^2_{AdS_3} = -(1+r^2)
\ dt^2+(1+r^2)^{-1} dr^2+r^2 d\phi^2
\\ \nn &&\equiv g_{tt} \ dt^2+g_{rr}
dr^2+g_{\phi\phi} d\phi^2
\\ \label{Sp} && ds^2_{S^3_+}
=\frac{1}{\cos^2\varphi}\left(d\theta_+^2+\sin^2\theta_+d\phi^2_{1+}
+\cos^2\theta_+d\phi^2_{2+}\right) \\ \nn &&\equiv
g_{\theta_+\theta_+}d\theta_+^2+g_{\phi_{1+}\phi_{1+}}d\phi^2_{1+}
+g_{\phi_{2+}\phi_{2+}}d\phi^2_{2+},
\\ \label{Sm} &&ds^2_{S^3_-}
=\frac{1}{\sin^2\varphi}\left(d\theta_-^2+\sin^2\theta_-d
\phi^2_{1-} +\cos^2\theta_-d \phi^2_{2-}\right)
\\ \nn &&\equiv
g_{\theta_-\theta_-}d\theta_-^2+g_{\phi_{1-}\phi_{1-}}d\phi^2_{1-}
+g_{\phi_{2-}\phi_{2-}}d\phi^2_{2-},
\\ \nn &&
ds^2_{S^1}=dw^2\equiv g_{ww}dw^2,
\\ \label{Bour} &&B= q r^2 \ d\phi \wedge dt
\\ \nn &&+\frac{q \sin^2 \theta_+}{\cos^2\varphi
\left(\cos^2\frac{\theta_+}{2}+\frac{\sin^2\frac{\theta_+}{2}}{\cos\varphi}\right)^2}\
d\phi_{1+} \wedge d\phi_{2+}
\\ \nn &&+\frac{q \sin^2 \theta_-}{\sin^2\varphi
\left(\cos^2\frac{\theta_-}{2}+\frac{\sin^2\frac{\theta_-}{2}}{\sin\varphi}\right)^2}\
d\phi_{1-} \wedge d\phi_{2-}\eea
\bea\nn &&\equiv b_{\phi t} \
d\phi \wedge dt +b_{\phi_{1+}\phi_{2+}}\ d\phi_{1+} \wedge
d\phi_{2+} +b_{\phi_{1-}\phi_{2-}}\ d\phi_{1-} \wedge
d\phi_{2-}.\eea

\setcounter{equation}{0}
%%%%%%%%%%%%%5%%%%%%%%%%%%%%%%%%%%%%%%%%%%%%%%%%%%%%
\section{The approach}
%%%%%%%%%%%%%5%%%%%%%%%%%%%%%%%%%%%%%%%%%%%%%%%%%%%%

Here, we will use the Polyakov type action for the bosonic string
in a $D$-dimensional curved space-time with metric tensor
$g_{MN}(x)$, interacting with a background 2-form gauge field
$b_{MN}(x)$ via Wess-Zumino term \bea\nn &&S^{P}=\int
d^{2}\xi\mathcal{L}^P,\h \mathcal{L}^P
=-\frac{1}{2}\left(T\sqrt{-\gamma}\gamma^{mn}G_{mn}-Q\varepsilon^{mn}
B_{mn}\right),\\ \nn && \xi^m=(\xi^0,\xi^1)=(\tau,\sigma),\h m,n =
0,1,\eea where  \bea\nn &&G_{mn}= \p_m X^M\p_n X^N g_{MN},\h
B_{mn}=\p_{m}X^{M}\p_{n}X^{N} b_{MN}, \\ \nn &&(\p_m=\p/\p\xi^m,\h
M,N = 0,1,\ldots,D-1),\eea are the fields induced on the string
worldsheet, $\gamma$ is the determinant of the auxiliary
worldsheet metric $\gamma_{mn}$, and $\gamma^{mn}$ is its inverse.
The position of the string in the background space-time is given
by $x^M=X^M(\xi^m)$, and $T=1/2\pi\alpha'$, $Q$ are the string
tension and charge, respectively. If we consider the action $S^P$
as a bosonic part of a supersymmetric one, we have to put $Q=\pm
T$. In what follows, $Q =T$.

The equations of motion for $X^M$ following from $S^P$ are: \bea
\nn
&&-g_{LK}\left[\p_m\left(\sqrt{-\gamma}\gamma^{mn}\p_nX^K\right) +
\sqrt{-\gamma}\gamma^{mn}\Gamma^K_{MN}\p_m X^M\p_n X^N\right]\\
\label{em} &&=\frac{1}{2}H_{LMN}\epsilon^{mn}\p_m X^M\p_n X^N,\eea
where ($\p_M=\p/\p x^M$) \bea\nn
&&\Gamma_{L,MN}=g_{LK}\Gamma^K_{MN}=\frac{1}{2}\left(\p_Mg_{NL}
+\p_Ng_{ML}-\p_Lg_{MN}\right),\\ \nn &&H_{LMN}= \p_L b_{MN}+ \p_M
b_{NL} + \p_N b_{LM},\eea are the components of the symmetric
connection corresponding to the metric $g_{MN}$, and the field
strength of the gauge field $b_{MN}$ respectively. The constraints
are obtained by varying the action $S^P$ with respect to
$\gamma_{mn}$: \bea\label{oc} \delta_{\gamma_{mn}}S^P=0\Rightarrow
\left(\gamma^{kl}\gamma^{mn}-2\gamma^{km}\gamma^{ln}\right)G_{mn}=0.\eea

Further on, we will  use {\it conformal gauge}
$\gamma^{mn}=\eta^{mn}=diag(-1,1)$ in which the string Lagrangian,
the Virasoro constraints and the equations of motion take the
following form: \bea\label{CG} &&\mathcal{L}
=\frac{T}{2}\left(G_{00}-G_{11}+2 B_{01}\right),\\ \nn
&&G_{00}+G_{11}=0,\h G_{01}=0, \\ \nn &&
g_{LK}\left[\left(\p_0^2-\p_1^2\right)X^K+ \Gamma_{MN}^K\left(\p_0
X^M \p_0 X^N-\p_1 X^M \p_1
X^N\right)\right]=H_{LMN}\p_0X^M\p_1X^N.\eea

Now, we {\it suppose} that there exist some number of commuting
Killing vector fields along part of $X^M$ coordinates and split
$X^M$ into two parts \bea\nn X^M=(X^\mu,X^a),\eea where $X^\mu$
are the isometric coordinates, while $X^a$ are the non-isometric
ones. The existence of isometric coordinates leads to the
following conditions on the background fields: \bea\label{cbf}
\p_\mu g_{MN}=0,\h \p_\mu b_{MN}=0.\eea Then from the string
action, we can compute the conserved charges \bea\label{CC}
Q_\mu=\int d\sigma \frac{\p \mathcal{L}}{\p(\p_0 X^\mu)}\eea under
the above conditions.

Next, we introduce the following ansatz for the string embedding
\bea\label{A} X^{\mu}(\tau,\sigma)=\Lambda^{\mu}\tau
+\tilde{X}^{\mu}(\alpha\sigma+\beta\tau),\h
X^{a}(\tau,\sigma)=\tilde{X}^{a}(\alpha\sigma+\beta\tau),\eea
where $\Lambda^{\mu}$, $\alpha$, $\beta$ are arbitrary parameters.
Further on, we will use the notation $\xi=\alpha\sigma+\beta\tau
$. Applying this ansatz, one can find that the equalities
(\ref{CG}), (\ref{CC}) become \bea\label{LA} \mathcal{L}=
\frac{T}{2}\Big[-(\alpha^2-\beta^2)g_{MN}\frac{d\tilde{X}^M}{d\xi}\frac{d\tilde{X}^N}{d\xi}
+2\Lambda^\mu\left(\beta g_{\mu N}+\alpha b_{\mu N}\right)
\frac{d\tilde{X}^N}{d\xi} +\Lambda^\mu\Lambda^\nu g_{\mu
\nu}\Big],\eea

\bea\label{V1}
G_{00}+G_{11}=(\alpha^2+\beta^2)g_{MN}\frac{d\tilde{X}^M}{d\xi}\frac{d\tilde{X}^N}{d\xi}
+2\beta\Lambda^\mu g_{\mu N} \frac{d\tilde{X}^N}{d\xi}
+\Lambda^\mu\Lambda^\nu g_{\mu \nu} =0,\eea

\bea\label{V2}  G_{01}&=&\alpha\beta
g_{MN}\frac{d\tilde{X}^M}{d\xi}\frac{d\tilde{X}^N}{d\xi}+\alpha\Lambda^\mu
g_{\mu N}\frac{d\tilde{X}^N}{d\xi} =0,\eea

\bea\nn
&&-(\alpha^2-\beta^2)\left[g_{LK}\frac{d^2\tilde{X}^K}{d\xi^2}+
\Gamma_{L,MN}\frac{d\tilde{X}^M}{d\xi}\frac{d\tilde{X}^N}{d\xi}\right]
+2\beta\Lambda^\mu\Gamma_{L,\mu N}\frac{d\tilde{X}^N}{d\xi}
+\Lambda^\mu\Lambda^\nu \Gamma_{L,\mu\nu}
\\ \label{EM}
&&= \alpha\Lambda^\mu H_{L\mu N}\frac{d\tilde{X}^N}{d\xi},\eea

\bea\label{Q} &&Q_{\mu}= \frac{T}{\alpha}\int
d\xi\left[\left(\beta g_{\mu N}+\alpha b_{\mu
N}\right)\frac{d\tilde{X}^N}{d\xi}+\Lambda^\nu g_{\mu
\nu}\right].\eea

Our next task is to try to solve the equations of motion
(\ref{EM}) for the isometric coordinates, i.e. for $L=\lambda$.
Due to the conditions (\ref{cbf}) imposed on the background
fields, we obtain that \bea\nn &&\Gamma_{\lambda,a
b}=\frac{1}{2}\left(\p_a g_{b \lambda}+\p_b
 g_{a \lambda}\right),\h  \Gamma_{\lambda,\mu a}=\frac{1}{2}\p_a g_{\mu\lambda}\h
 \Gamma_{\lambda,\mu \nu}=0,
 \\ \nn && H_{\lambda a b}=\p_a b_{b \lambda}+\p_b b_{\lambda a}, \h
 H_{\lambda \mu a}=\p_a b_{\lambda\mu}, \h  H_{\lambda \mu \nu}=0.
 \eea
By using this, one can find the following first integrals for
$\tilde{X}^\mu$: \bea\label{FIM} \frac{d\tilde{X}^{\mu}}{d\xi}=
\frac{1}{\alpha^2-\beta^2}\left[g^{\mu\nu}
\left(C_\nu-\alpha\Lambda^\rho
b_{\nu\rho}\right)+\beta\Lambda^\mu\right] -g^{\mu\nu}g_{\nu
a}\frac{d\tilde{X}^{a}}{d\xi},\eea where $C_\nu$ are arbitrary
integration constants. Therefore, according to our ansatz
(\ref{A}), the solutions for the string coordinates $X^\mu$ can be
written as \bea\label{MS} X^{\mu}(\tau,\sigma)=\Lambda^{\mu}\tau
+\frac{1}{\alpha^2-\beta^2}\int d\xi \left[g^{\mu\nu}
\left(C_\nu-\alpha\Lambda^\rho
b_{\nu\rho}\right)+\beta\Lambda^\mu\right] -\int g^{\mu\nu}g_{\nu
a}d\tilde{X}^{a}(\xi).\eea

Now, let us turn to the remaining equations of motion
corresponding to $L=a$, where \bea\nn &&\Gamma_{a,\mu
b}=-\frac{1}{2}(\p_a g_{b \mu}-\p_b g_{a \mu}),\h \Gamma_{a,\mu
\nu}=-\frac{1}{2}\p_a g_{\mu\nu},
\\ \nn &&H_{a \mu\nu}=\p_a b_{\mu\nu},\h H_{a \mu b}=-\p_a b_{b \mu}+\p_b b_{a \mu}.\eea
Taking this into account and replacing the first integrals for
$\tilde{X}^{\mu}$ already found, one can write these equations in
the form (prime is used for $d/d\xi$) \bea\label{Ea}
(\alpha^2-\beta^2)\left[h_{a
b}\tilde{X}^{b''}+\Gamma^{h}_{a,bc}\tilde{X}^{b'}\tilde{X}^{c'}\right]
= 2\p_{[a} A_{b]}\tilde{X}^{b'}-\p_a U, \eea where \bea\label{Ma}
&&h_{a b}= g_{a b}-g_{a \mu}g^{\mu\nu}g_{\nu b},\h
\Gamma^{h}_{a,bc}=\frac{1}{2}\left(\p_b h_{ca}+\p_c h_{ba}-\p_a
h_{bc}\right)
\\ \label{VPa} &&A_a= g_{a\mu}g^{\mu\nu}
\left(C_\nu-\alpha\Lambda^\rho
b_{\nu\rho}\right)+\alpha\Lambda^\mu b_{a\mu},
\\ \label{SP} &&U=
\frac{1/2}{\alpha^2-\beta^2}\left[\left(C_\mu-\alpha\Lambda^\rho
b_{\mu\rho}\right)g^{\mu\nu} \left(C_\nu-\alpha\Lambda^\lambda
b_{\nu\lambda}\right)+\alpha^2\Lambda^\mu\Lambda^\nu
g_{\mu\nu}\right].\eea

The Virasoro constraints (\ref{V1}), (\ref{V2}) become:
\bea\label{V12} \frac{1}{2}(\alpha^2-\beta^2) h_{a
b}\tilde{X}^{a'}\tilde{X}^{b'}+U=0,\h \alpha\Lambda^\mu C_\mu
=0.\eea

Finally, let us write down the expressions for the conserved
charges (\ref{Q}) \bea\nn Q_\mu &=&\frac{T}{\alpha^2-\beta^2}\int
d\xi \left[\frac{\beta}{\alpha}C_\mu+\alpha\Lambda^\nu g_{\mu\nu}
+ b_{\mu\nu}g^{\nu\rho} \left(C_\rho-\alpha\Lambda^\lambda
b_{\rho\lambda}\right)\right. \\ \label{Qmu}
&+&\left.(\alpha^2-\beta^2) \left(b_{\mu
a}-b_{\mu\nu}g^{\nu\rho}g_{\rho
a}\right)\tilde{X}^{a'}\right].\eea

\setcounter{equation}{0}
%%%%%%%%%%%%%5%%%%%%%%%%%%%%%%%%%%%%%%%%%%%%%%%%%%%%
\section{String solutions in $AdS_3\times S^3\times S^3\times
S^1$ with $B$-field}
%%%%%%%%%%%%%5%%%%%%%%%%%%%%%%%%%%%%%%%%%%%%%%%%%%%%

In accordance with our notations \bea\nn
&&\mu=\left(t,\phi,\phi_{1+},\phi_{2+},\phi_{1-},\phi_{2-},w\right),\h
a=\left(r,\theta_{+},\theta_{-}\right),
\\ \nn &&g_{\mu\nu}=\left(g_{t t},g_{\phi
\phi},g_{\phi_{1+}\phi_{1+}},g_{\phi_{2+}\phi_{2+}},
g_{\phi_{1-}\phi_{1-}},g_{\phi_{2-}\phi_{2-}},g_{ww}\right),\\ \nn
&&g_{ab}=\left(g_{r
r},g_{\theta_{+}\theta_{+}},g_{\theta_{-}\theta_{-}}\right),
\\ \nn
&&g_{a \mu}=0,\h h_{ab}=g_{ab},
\\ \nn &&b_{\mu\nu}=(b_{\phi t},b_{\phi_{1+}\phi_{2+}},b_{\phi_{1-}\phi_{2-}}),\h b_{a \nu}=0,
\\ \label{idc} &&A_a=0,\eea

where

\bea\nn &&g_{t t}=-(1+r^2), \h g_{r r}=(1+r^2)^{-1}, \h g_{\phi
\phi}=r^2,
\\ \nn &&g_{\theta_{+}\theta_{+}}=\frac{1}{\cos^2\varphi},
\h g_{\phi_{1+} \phi_{1+}}
=\frac{1}{\cos^2\varphi}\sin^2\theta_{+},\h g_{\phi_{2+}
\phi_{2+}}=\frac{1}{\cos^2\varphi}\cos^2\theta_{+},
\\ \nn &&g_{\theta_{-}\theta_{-}}=\frac{1}{\sin^2\varphi},
\h g_{\phi_{1-} \phi_{1-}}
=\frac{1}{\sin^2\varphi}\sin^2\theta_{-},
\h g_{\phi_{2-} \phi_{2-}}=\frac{1}{\sin^2\varphi}\cos^2\theta_{-}, \\
\nn &&g_{ww}=1,
\\ \label{bfs} &&b_{t\phi}=-q r^2,\\ \nn
&&b_{\phi_{1+}\phi_{2+}}=\frac{q \sin^2 \theta_+}{\cos^2\varphi
\left(\cos^2\frac{\theta_+}{2}+\frac{\sin^2\frac{\theta_+}{2}}{\cos\varphi}\right)^2},\\
\nn &&b_{\phi_{1-}\phi_{2-}}=\frac{q \sin^2
\theta_-}{\sin^2\varphi
\left(\cos^2\frac{\theta_-}{2}+\frac{\sin^2\frac{\theta_-}{2}}{\sin\varphi}\right)^2}.\eea

The effective scalar potential (\ref{SP}) can be computed to be
\bea\label{sp} U=\sum_{i=1}^{4}U_i,\eea where

\bea\label{U1} U_1(r)&=& \frac{1/2}{\alpha^{2}-\beta^{2}}
\Bigg\{-\frac{(C_t+\alpha\Lambda^{\phi}qr^{2})^{2}}{1+r^{2}}
+\frac{(C_{\phi}-\alpha\Lambda^{t}qr^{2})^{2}}{r^{2}}
\\ \nn &&-\alpha^{2}\left[\left(\Lambda^{t}\right)^{2}(1+r^2)
-\left(\Lambda^{\phi}\right)^{2}r^2\right]\Bigg\},\eea

\bea\label{U2} U_2(\theta_{+})&=& \frac{1/2}{\alpha^{2}-\beta^{2}}
\Bigg\{\left[C_{\phi_{1+}}-\frac{\alpha\Lambda^{\phi_{2+}}
q\sin^2\theta_+}{\cos^2\varphi\left(\cos^2\frac{\theta_+}{2}
+\frac{\sin^2\frac{\theta_+}{2}}{\cos\varphi}\right)^{2}}
\right]^{2} \frac{\cos^2\varphi}{\sin^2\theta_+} \\ \nn
&&+\left[C_{\phi_{2+}}+\frac{\alpha\Lambda^{\phi_{1+}}
q\sin^2\theta_+}{\cos^2\varphi\left(\cos^2\frac{\theta_+}{2}
+\frac{\sin^2\frac{\theta_+}{2}}{\cos\varphi}\right)^{2}}
\right]^{2} \frac{\cos^2\varphi}{\cos^2\theta_+}
\\ \nn &&+\frac{\alpha^{2}}{\cos^2\varphi}
\left[\left(\Lambda^{\phi_{1+}}\right)^{2}\sin^2\theta_+
+\left(\Lambda^{\phi_{2+}}\right)^{2}
\cos^2\theta_+\right]\Bigg\},\eea

\bea\label{U3} U_3(\theta_{-})&=& \frac{1/2}{\alpha^{2}-\beta^{2}}
\Bigg\{\left[C_{\phi_{1-}}-\frac{\alpha\Lambda^{\phi_{2-}}
q\sin^2\theta_-}{\sin^2\varphi\left(\cos^2\frac{\theta_-}{2}
+\frac{\sin^2\frac{\theta_-}{2}}{\sin\varphi}\right)^{2}}
\right]^{2} \frac{\sin^2\varphi}{\sin^2\theta_-} \\ \nn
&&+\left[C_{\phi_{2-}}+\frac{\alpha\Lambda^{\phi_{1-}}
q\sin^2\theta_-}{\sin^2\varphi\left(\cos^2\frac{\theta_-}{2}
+\frac{\sin^2\frac{\theta_-}{2}}{\sin\varphi}\right)^{2}}
\right]^{2} \frac{\sin^2\varphi}{\cos^2\theta_-}
\\ \nn &&+\frac{\alpha^{2}}{\sin^2\varphi}
\left[\left(\Lambda^{\phi_{1-}}\right)^{2}\sin^2\theta_-
+\left(\Lambda^{\phi_{2-}}\right)^{2} \cos^2\theta_-\right]
\Bigg\} ,\eea

\bea\label{U4} U_4=C_w^2+(\Lambda^w)^{2}=const .\eea

By using that in the case under consideration the metric $g_{ab}$
is diagonal, one can prove that the equations of motion (\ref{Ea})
possess the following first integrals \bea\label{fia}
\frac{d\tilde{X}^a}{d\xi}=\sqrt{\frac{C_a-2
U_a}{(\alpha^2-\beta^2)g_{aa}}},\eea where
$\tilde{X}^a=(r,\theta_{+},\theta_{-})$,
$C_a=(C_{r},C_{\theta_{+}},C_{\theta_{-}})$ are arbitrary
integration constants and $U_a=(
U_1(r),U_2(\theta_{+}),U_3(\theta_{-}))$.

The replacement of (\ref{fia}) into (\ref{V12}) reduces the first
Virasoro constraint to the following equality \bea\label{Vir1r}
C_{r}+C_{\theta_{+}}+C_{\theta_{-}}=0.\eea Thus, the Virasoro
constraints are simplified to relations between the integration
constants and embedding parameters on this type of string
solutions.

%%%%%%%%%%%%%5%%%%%%%%%%%%%%%%%%%%%%%%%%%%%%%%%%%%%%
\subsection{Solutions in $AdS_3$}
%%%%%%%%%%%%%5%%%%%%%%%%%%%%%%%%%%%%%%%%%%%%%%%%%%%%
For the $AdS_3$ subspace, (\ref{fia}) gives \bea\nn d\xi=
\frac{dr}{\sqrt{\frac{(C_{r}-2U_1(r))(1+r^2)}{\alpha^{2}-\beta^{2}}}}.\eea
By using the expression (\ref{U1}) for $U_1(r)$ and introducing
the variable $y=r^2$, one obtains \bea\label{dxi} d\xi=
\frac{\alpha^{2}-\beta^{2}}{2 \alpha
\sqrt{(1-q^2)\left[\left(\Lambda^\phi\right)^{2}-\left(\Lambda^t\right)^{2}\right]}}
\frac{dy}{\sqrt{(y_p-y)(y-y_m)(y-y_n)}},\eea where \bea\nn y_p>y>
y_m \geq 0,\h y_n<0, \eea and $y_p$, $y_m$, $y_n$ satisfy the
equalities \bea\nn
&&y_p+y_m+y_n=\frac{1}{\alpha^{2}(1-q^2)\left[\left(\Lambda^\phi\right)^2-\left(\Lambda^t\right)^2\right]}
\\ \nn
&&\left[C_r(\alpha^2-\beta^2)-\alpha\left(\alpha\left(\Lambda^\phi\right)^2-2\alpha\left(\Lambda^t\right)^2
-2q\left(C_\phi\Lambda^t+C_t\Lambda^\phi\right)+q^2\alpha\left(\Lambda^t\right)^2\right)\right],
\\ \label{ypm} && y_p y_m+y_p y_n+y_m y_n = -\frac{1}{\alpha^{2}(1-q^2)\left[\left(\Lambda^\phi\right)^2-\left(\Lambda^t\right)^2\right]}
\\ \nn &&\left[C_r(\alpha^2-\beta^2)+C_t^2-C_\phi^2+\alpha^2\left(\Lambda^t\right)^2+2q\alpha C_\phi\Lambda^t\right],
\\ \nn && y_p y_m y_n=- \frac{C_\phi^2}{\alpha^{2}(1-q^2)\left[\left(\Lambda^\phi\right)^2-\left(\Lambda^t\right)^2\right]}.\eea

Integrating (\ref{dxi}) and inverting \bea\nn \xi(y)=
\frac{\alpha^2-\beta^2}{\alpha
\sqrt{(1-q^2)\left[\left(\Lambda^\phi\right)^2-\left(\Lambda^t\right)^2\right](y_p-y_n)}}\
\mathbf{F}\left(\arcsin\sqrt{\frac{y_p-y}{y_p-y_m}},\frac{y_p-y_m}{y_p-y_n}\right)\eea
to $y(\xi)$, one finds the following solution \bea\label{yxi}
y(\xi)=(y_p-y_n)\ {\mathbf{DN}}^2\left[\frac{\alpha
\sqrt{(1-q^2)\left[\left(\Lambda^\phi\right)^2-\left(\Lambda^t\right)^2\right](y_p-y_n)}}
{\alpha^2-\beta^2}\ \xi,\frac{y_p-y_m}{y_p-y_n}\right]+y_n,\eea
where $\mathbf{F}$ is the incomplete elliptic integral of first
kind and $\mathbf{DN}$ is one of the Jacobi elliptic functions.

Now we are going to find the solutions for the isometric
coordinates $t(\xi)$ and $\phi(\xi)$. In accordance with
(\ref{FIM}), the first integrals for $\tilde{X}^t$ and
$\tilde{X}^{\phi}$ can be computed to be given by \bea\nn
&&\frac{d\tilde{X}^t}{d\xi}= \frac{1}{\alpha^2-\beta^2}
\left[\beta\Lambda^t -q\alpha\Lambda^\phi
-\frac{C_t-q\alpha\Lambda^\phi}{1+y}\right],
\\ \nn &&\frac{d\tilde{X}^\phi}{d\xi}= \frac{1}{\alpha^2-\beta^2}
\left(\beta\Lambda^\phi-q\alpha\Lambda^t+\frac{C_\phi}{y}\right).\eea

Integrating and using (\ref{dxi}), we obtain \bea\label{tsol}
&&t(\tau,\xi)=\Lambda^t\tau+\frac{1}{\alpha\sqrt{(1-q^2)\left[\left(\Lambda^\phi\right)^2
-\left(\Lambda^t\right)^2\right](y_p-y_n)}}
\\ \nn &&\left[\left(\beta \Lambda^t-q \alpha\Lambda^\phi\right)\
\mathbf{F}\left(\arcsin\sqrt{\frac{y_p-y}{y_p-y_m}},\frac{y_p-y_m}{y_p-y_n}\right)\right.
\\ \nn &&-\left. \frac{C_t-q \alpha\Lambda^\phi}{1+y_p}
\ \mathbf{\Pi}\left(\arcsin\sqrt{\frac{y_p-y}{y_p-y_m}},
\frac{y_p-y_m}{1+y_p},\frac{y_p-y_m}{y_p-y_n}\right)\right],
\\ \label{Ff} &&\phi(\tau,\xi) =\Lambda^\phi\tau + \frac{1}{\alpha\sqrt{(1-q^2)\left[\left(\Lambda^\phi\right)^2
-\left(\Lambda^t\right)^2\right](y_p-y_n)}}
\\ \nn &&\left[\left(\beta \Lambda^\phi-q\alpha\Lambda^t\right)\
\mathbf{F}\left(\arcsin\sqrt{\frac{y_p-y}{y_p-y_m}},\frac{y_p-y_m}{y_p-y_n}\right)\right.
\\ \nn &&+\left. \frac{C_\phi}{y_p}
\ \mathbf{\Pi}\left(\arcsin\sqrt{\frac{y_p-y}{y_p-y_m}},
\frac{y_p-y_m}{y_p},\frac{y_p-y_m}{y_p-y_n}\right)\right],\eea
where $\mathbf{\Pi}$ is the incomplete elliptic integral of third
kind.

%%%%%%%%%%%%%5%%%%%%%%%%%%%%%%%%%%%%%%%%%%%%%%%%%%%%
\subsection{Solutions on the two $S^3$}
%%%%%%%%%%%%%5%%%%%%%%%%%%%%%%%%%%%%%%%%%%%%%%%%%%%%
It is clear from (\ref{Sp}) and (\ref{Sm}) that the solutions on
the two three-spheres $S^3_+$ and $S^3_-$ can be obtained from
each other by the exchanges $\sin\varphi \leftrightarrow
\cos\varphi$ and $+\leftrightarrow -$ in the subscripts of the
coordinates, the corresponding integration constants, and in the
superscripts of the embedding parameters. That is why we are going
to present here the string solutions for one of the spheres only,
say $S^3_-$.

The non-isometric coordinate on $S^3_-$ is $\theta_-$ for which
the first integral (\ref{fia}) reads \bea\label{fim}
\frac{d\theta_-}{d\xi}=\sqrt{\frac{\sin^2\varphi}{\alpha^2-\beta^2}
\left[C_{\theta_-}-2 U_3(\theta_-)\right]}.\eea $U_3(\theta_-)$ is
given in (\ref{U3}).

Now we introduce the variable \be\nn
\gamma=\cos^2\frac{\theta_-}{2}.\eea This allows us to rewrite
(\ref{fim}) in the following form \bea\label{dxi1} d\xi&=&
\frac{\alpha^{2}-\beta^{2}}{\sin\varphi} (-a_{10})^{-1/2}
\left(1-(1-\sin\varphi)\gamma\right)^{2}(2\gamma-1) \\ \nn &&
\left[(\gamma_1-\gamma)(\gamma-\gamma_2)(\gamma-\gamma_3)(\gamma-\gamma_4)(\gamma-\gamma_5)
(\gamma-\gamma_6)\right.
\\ \nn &&\left.(\gamma-\gamma_7)(\gamma-\gamma_8)(\gamma-\gamma_9)(\gamma-\gamma_{10})\right]^{-1/2}d\gamma
,\h \gamma_1=\gamma_{max} .\eea The corresponding computations are
given in an Appendix. Next, we integrate \bea\label{xi1} \xi&=&
\frac{\alpha^{2}-\beta^{2}}{\sin\varphi}
(-a_{10})^{-1/2}\int_{\gamma}^{\gamma_{max}}
\left(1-(1-\sin\varphi)u\right)^{2}(2u-1) \\ \nn &&
\left[(\gamma_{max}-u)(u-\gamma_2)(u-\gamma_3)(u-\gamma_4)(u-\gamma_5)
(u-\gamma_6)\right.
\\ \nn
&&\left.(u-\gamma_7)(u-\gamma_8)(u-\gamma_9)(u-\gamma_{10})\right]^{-1/2}du
,\eea and introduce new integration variable \bea\nn \delta=
\frac{\gamma_{max}-u}{\gamma_{max}-\gamma}.\eea Then (\ref{xi1})
becomes
\bea\label{xi2}  \xi(\gamma)&=&
\frac{\alpha^{2}-\beta^{2}}{\sin\varphi} (-a_{10})^{-1/2}
(1-\gamma_{max}(1-\sin\varphi))^{2}(2\gamma_{max}-1)(\gamma_{max}-\gamma)^{1/2}
\\ \nn &&\left[\prod_{i=2}^{10}(\gamma_{max}-\gamma_i)\right]^{-1/2}\int_{0}^{1}
\delta^{-1/2}\left(1+\frac{(\gamma_{max}-\gamma)(1-\sin\varphi)}{(1-\gamma_{max}(1-\sin\varphi))}\delta\right)^{2}
\\ \nn &&\left(1-\frac{2(\gamma_{max}-\gamma)}{2\gamma_{max}-1}\delta\right)
\prod_{i=2}^{10}\left(1-\frac{\gamma_{max}-\gamma}{\gamma_{max}-\gamma_i}\delta\right)^{-1/2}d\delta.\eea

Comparing the integral in (\ref{xi2}) with the integral
representation of the Lauricella hypergeometric functions of
 many variables $F_D^{(n)}$ \cite{PBMv3}
\bea\label{FDn} &&F_D^{(n)}(a;b_1,\ldots,b_n;c;z_1,\ldots,z_n)=
\\ \nn &&\frac{\Gamma(c)}{\Gamma(a)\Gamma(c-a)}\int_{0}^{1}
\delta^{a-1}(1-\delta)^{c-a-1}(1-z_1\delta)^{-b_1}\ldots
(1-z_n\delta)^{-b_n}d\delta, \\ \nn && Re (a)>0,\h Re (c-a)>0,\eea
one finds \bea\label{xisols} \xi(\gamma)&=&2
\frac{\alpha^{2}-\beta^{2}}{\sin\varphi} (-a_{10})^{-1/2}
(1-\gamma_{max}(1-\sin\varphi))^{2}(2\gamma_{max}-1)(\gamma_{max}-\gamma)^{1/2}
\\ \nn &&\left[\prod_{i=2}^{10}(\gamma_{max}-\gamma_i)\right]^{-1/2}
F_D^{(11)}(1/2;b_1,\ldots,b_{11};3/2;z_1,\ldots,z_{11}),\eea where
\bea\label{bz1} &&b_1=-2,\h
z_1=-\frac{(\gamma_{max}-\gamma)(1-\sin\varphi)}{1-\gamma_{max}(1-\sin\varphi)},
\\ \nn &&b_2=-1,\h z_2=\frac{2(\gamma_{max}-\gamma)}{2\gamma_{max}-1},
\\ \nn &&b_k=1/2,\h z_k=\frac{\gamma_{max}-\gamma}{\gamma_{max}-\gamma_{k-1}},\h k=3,\ldots,11.\eea
This is our final result for $\xi(\gamma)$. Unfortunately this
solution is not invertible, so we can not write down
$\gamma(\xi)$.

Now, let us proceed with obtaining the solutions for the isometric
coordinates on $S^3_-$. From (\ref{FIM}) one finds the following
first integrals: \bea\label{Xt1m}
\frac{d\tilde{X}^{\phi_{1-}}}{d\xi}=
\frac{1}{\alpha^{2}-\beta^{2}} \left\{\beta \Lambda^{\phi_{1-}}
+\sin^2\varphi
\left[\frac{C_{\phi_{1-}}}{\sin^2\theta_-}-\frac{q\alpha\Lambda^{\phi_{2-}}}{\sin^2\varphi
\left(\cos^2\frac{\theta_-}{2}+\frac{\sin^2\frac{\theta_-}{2}}{\sin\varphi}\right)^2}\right]\right\},\eea

\bea\label{Xt2m} \frac{d\tilde{X}^{\phi_{2-}}}{d\xi}=
\frac{1}{\alpha^{2}-\beta^{2}} \left\{\beta \Lambda^{\phi_{2-}}
+\frac{\sin^2\varphi}{\cos^2\theta_-}
\left[C_{\phi_{2-}}+\frac{q\alpha\Lambda^{\phi_{1-}}\sin^2\theta_-}{\sin^2\varphi
\left(\cos^2\frac{\theta_-}{2}+\frac{\sin^2\frac{\theta_-}{2}}{\sin\varphi}\right)^2}\right]\right\}.\eea

Introducing the variable $\gamma$ and using (\ref{dxi1}), one
arrives at \bea\label{phit1m} \tilde{X}^{\phi_{1-}} \equiv
\tilde{\phi}_{1-}&=& 2\sin\varphi  (-a_{10})^{-1/2}
(1-\gamma_{max}(1-\sin\varphi))^{2}(2\gamma_{max}-1)(\gamma_{max}-\gamma)^{1/2}
\\ \nn &&\left[\prod_{i=2}^{10}(\gamma_{max}-\gamma_i)\right]^{-1/2}
\left[\frac{\beta\Lambda^{\phi_{1-}}}{\sin^{2}\varphi}
F_D^{(11)}(1/2;b_1,\ldots,b_{11};3/2;z_1,\ldots,z_{11})\right.
\\ \nn &&+\left. \frac{C_{\phi_{1-}}}{4(1-\gamma_{max})\gamma_{max}}
F_D^{(13)}(1/2;b_1,\ldots,b_{11},b_{12},b_{13};3/2;z_1,\ldots,z_{11},z_{12},z_{13})\right.
\\ \nn &&\left.-q\alpha\Lambda^{\phi_{2-}}(1-\gamma_{max}(1-\sin\varphi))^{-2}
F_D^{(10)}(1/2;b_2,\ldots,b_{11};3/2;z_2,\ldots,z_{11})\right]
,\eea

\bea\label{phit2m}\tilde{X}^{\phi_{2-}} \equiv
\tilde{\phi}_{2-}&=& 2\sin\varphi  (-a_{10})^{-1/2}
(1-\gamma_{max}(1-\sin\varphi))^{2}(2\gamma_{max}-1)(\gamma_{max}-\gamma)^{1/2}
\\ \nn &&\left[\prod_{i=2}^{10}(\gamma_{max}-\gamma_i)\right]^{-1/2}
\left[\frac{\beta\Lambda^{\phi_{2-}}}{\sin^{2}\varphi}
F_D^{(11)}(1/2;b_1,\ldots,b_{11};3/2;z_1,\ldots,z_{11})\right.
\\ \nn &&\left.+\frac{C_{\phi_{2-}}}{(2\gamma_{max}-1)^{2}}
F_D^{(12)}(1/2;b_1,\ldots,b_{11},c_{12};3/2;z_1,\ldots,z_{11},y_{12})\right.
\\ \nn &&\left.+4q\alpha\Lambda^{\phi_{1-}}(2\gamma_{max}-1)^{-2}
\gamma_{max}(1-\gamma_{max})(1-\gamma_{max}(1-\sin\varphi))^{-2}\times\right.
\\ \nn &&\left.F_D^{(15)}(1/2;b_1,\ldots,b_{11},c_{12},c_{13},c_{14},c_{15};3/2;
z_1,\ldots,z_{11},y_{12},z_{12},z_{13},y_{15})\right],\eea where
\bea\label{bz2} &&b_{12}=1,\h
z_{12}=-\frac{\gamma_{max}-\gamma}{1-\gamma_{max}},
\\ \nn &&b_{13}=1,\h
z_{13}=\frac{\gamma_{max}-\gamma}{\gamma_{max}},
\\ \nn &&c_{12}=2,\h y_{12}=2\frac{\gamma_{max}-\gamma}{2\gamma_{max}-1},
\\ \nn &&c_{13}=-1,\h c_{14}=-1,
\\ \nn &&c_{15}=2,\h y_{15}=-\frac{(\gamma_{max}-\gamma)(1-\sin\varphi)}{1-\gamma_{max}(1-\sin\varphi)}.\eea

Therefore, according to (\ref{MS}), the solutions for the
isometric coordinates on $S^3_-$ are given by \bea\nn
\phi_{1-}(\tau,\xi)=\Lambda^{\phi_{1-}}\tau+\tilde{\phi}_{1-}(\xi),\h
\phi_{2-}(\tau,\xi)=\Lambda^{\phi_{2-}}\tau+\tilde{\phi}_{2-}(\xi).\eea

\setcounter{equation}{0}
%%%%%%%%%%%%%5%%%%%%%%%%%%%%%%%%%%%%%%%%%%%%%%%%%%%%
\section{Conserved charges}
%%%%%%%%%%%%%5%%%%%%%%%%%%%%%%%%%%%%%%%%%%%%%%%%%%%%

The expressions for the conserved charges corresponding to the
isometric coordinates can be found from (\ref{Qmu}) to be
\bea\label{Es} Q_t\equiv -E_s= \frac{T}{\alpha^{2}-\beta^{2}} \int
\left[\frac{\beta}{\alpha}C_t-\alpha\Lambda^t(1+r^2)
-q\left(C_\phi-q\alpha\Lambda^t r^2\right)\right] d\xi,\eea

\bea\label{S} Q_\phi\equiv S= \frac{T}{\alpha^{2}-\beta^{2}} \int
\left[\frac{\beta}{\alpha}C_\phi-\alpha\Lambda^\phi
r^2-q\frac{r^2}{1+r^2} \left(C_t+q\alpha\Lambda^\phi
r^2\right)\right]d\xi,\eea

\bea\label{J1p} &&Q_{\phi_{1+}}\equiv J_{1+}=
\frac{T}{\alpha^{2}-\beta^{2}}
\int\left[\frac{\beta}{\alpha}C_{\phi_{1+}}+\frac{\alpha\Lambda^{\phi_{1+}}}{\cos^2\varphi}
\sin^2\theta_+ \right.
\\ \nn &&\left. +\frac{q\sin^2\theta_+}{\cos^2\theta_+\left(\cos^2\frac{\theta_+}{2}
+\frac{\sin^2\frac{\theta_+}{2}}{\cos\varphi}\right)^2}
\left(C_{\phi_{2+}}+\frac{q\alpha\Lambda^{\phi_{1+}}\sin^2\theta_+}{\cos^2\varphi
\left(\cos^2\frac{\theta_+}{2}
+\frac{\sin^2\frac{\theta_+}{2}}{\cos\varphi}\right)^2}\right)\right]
d\xi,\eea

\bea\label{J2p} &&Q_{\phi_{2+}}\equiv J_{2+}=
\frac{T}{\alpha^{2}-\beta^{2}}
\int\left[\frac{\beta}{\alpha}C_{\phi_{2+}}+\frac{\alpha\Lambda^{\phi_{2+}}}{\cos^2\varphi}
\cos^2\theta_+\right.
\\ \nn  &&
\left.-\frac{q}{\left(\cos^2\frac{\theta_+}{2}
+\frac{\sin^2\frac{\theta_+}{2}}{\cos\varphi}\right)^2}
\left(C_{\phi_{1+}}-\frac{q\alpha\Lambda^{\phi_{2+}}\sin^2\theta_+}{\cos^2\varphi
\left(\cos^2\frac{\theta_+}{2}
+\frac{\sin^2\frac{\theta_+}{2}}{\cos\varphi}\right)^2}\right)\right]
d\xi,\eea

\bea\label{J1m} &&Q_{\phi_{1-}}\equiv J_{1-}=
\frac{T}{\alpha^{2}-\beta^{2}}
\int\left[\frac{\beta}{\alpha}C_{\phi_{1-}}+\frac{\alpha\Lambda^{\phi_{1-}}}{\sin^2\varphi}
\sin^2\theta_-\right.
\\ \nn &&\left.+\frac{q\sin^2\theta_-}{\cos^2\theta_-\left(\cos^2\frac{\theta_-}{2}
+\frac{\sin^2\frac{\theta_-}{2}}{\sin\varphi}\right)^2}
\left(C_{\phi_{2-}}+\frac{q\alpha\Lambda^{\phi_{1-}}\sin^2\theta_-}{\sin^2\varphi
\left(\cos^2\frac{\theta_-}{2}
+\frac{\sin^2\frac{\theta_-}{2}}{\sin\varphi}\right)^2}\right)\right]
d\xi,\eea

\bea\label{J2m} &&Q_{\phi_{2-}}\equiv J_{2-}=
\frac{T}{\alpha^{2}-\beta^{2}}
\int\left[\frac{\beta}{\alpha}C_{\phi_{2-}}+\frac{\alpha\Lambda^{\phi_{2-}}}{\sin^2\varphi}
\cos^2\theta_-\right. \\ \nn
&&\left.-\frac{q}{\left(\cos^2\frac{\theta_-}{2}
+\frac{\sin^2\frac{\theta_-}{2}}{\sin\varphi}\right)^2}
\left(C_{\phi_{1-}}-\frac{q\alpha\Lambda^{\phi_{2-}}\sin^2\theta_-}{\sin^2\varphi
\left(\cos^2\frac{\theta_-}{2}
+\frac{\sin^2\frac{\theta_-}{2}}{\sin\varphi}\right)^2}\right)\right]
d\xi.\eea

Here we introduced the following notations: $E_s$ is the string
energy, $S$ is the spin of the string in $AdS_3$, $J_{1+}$,
$J_{2+}$, $J_{1-}$, $J_{2-}$, are the angular momenta on the two
three spheres $S^3_{\pm}$.

In order to compute $E_s$ and $S$, we introduce the variable
$y=r^2$ and use the expression (\ref{dxi}) for $d\xi$ in $AdS_3$.
This leads to the following results \bea\label{Ess} &&E_s= \frac{2
T}{\sqrt{(1-q^2)\left[\left(\Lambda^\phi\right)^2
-\left(\Lambda^t\right)^2\right](y_p-y_n)}}
\\ \nn &&\left[\left(\Lambda^t-\frac{\beta}{\alpha^2}C_t+q\frac{C_\phi}{\alpha}\right)
\mathbf{K}\left(1-\frac{y_m-y_n}{y_p-y_n}\right) +\right.
\\ \nn &&\left. (1-q^2)\Lambda^t \left(y_n \ \mathbf{K}\left(1-\frac{y_m-y_n}{y_p-y_n}\right)
+(y_p-y_n)\
\mathbf{E}\left(1-\frac{y_m-y_n}{y_p-y_n}\right)\right)\right],
\\ \label{Ss} &&S= \frac{2 T}{\sqrt{(1-q^2)\left[\left(\Lambda^\phi\right)^2
-\left(\Lambda^t\right)^2\right](y_p-y_n)}}
\\ \nn &&\left[\left(\frac{\beta}{\alpha^2}C_\phi-q\frac{C_t}{\alpha}+\Lambda^\phi q^2\right)
\mathbf{K}\left(1-\frac{y_m-y_n}{y_p-y_n}\right) +\right.
\\ \nn &&\left. (1-q^2)\Lambda^\phi \left(y_n \ \mathbf{K}\left(1-\frac{y_m-y_n}{y_p-y_n}\right)
+(y_p-y_n)\
\mathbf{E}\left(1-\frac{y_m-y_n}{y_p-y_n}\right)\right)\right.
\\ \nn &&\left. +\frac{q\frac{C_t}{\alpha}-q^2\Lambda^\phi}{1+y_p}\
\mathbf{\Pi}\left(\frac{y_p-y_m}{1+y_p},1-\frac{y_m-y_n}{y_p-y_n}\right)\right],\eea
where $\mathbf{K}$, $\mathbf{E}$ and $\mathbf{\Pi}$ are the
complete elliptic integrals of first, second and third kind.

To find the final expressions for the angular momenta, we now
introduce the variable $\gamma$ ($\gamma=\cos^2\frac{\theta_+}{2}$
for $S^3_+$ and $\gamma=\cos^2\frac{\theta_-}{2}$ for $S^3_-$).
Here we present the results for $J_{1-}$ and $J_{2-}$ only. By
using (\ref{dxi1}) we derive: \bea\label{J1mf} &&J_{1-}=\frac{2\pi
T}{\sin\varphi}(-a_{10})^{-1/2}
(1-\gamma_{max}(1-\sin\varphi))^{2}(2\gamma_{max}-1)(\gamma_{max}-\gamma_{min})^{1/2}
\\ \nn
&&\left[\prod_{i=2}^{10}(\gamma_{max}-\gamma_i)\right]^{-1/2}\times
\\ \nn &&
\left[\frac{\beta}{\alpha}C_{\phi_{1-}}
F_D^{(10)}(1/2;b_1,\ldots,b_{k-1},b_{k+1},\ldots,b_{11};1;Z_1,\ldots,Z_{k-1},Z_{k+1},\ldots,Z_{11})
\right.
\\ \nn && \left.
+\frac{4\alpha\Lambda^{\phi_{1-}}}{\sin^2\varphi}\gamma_{max}(1-\gamma_{max})\times\right.
\\ \nn && \left. F_D^{(12)}(1/2;b_1,\ldots,b_{k-1},b_{k+1},\ldots,b_{11},b_{12},b_{13};1;Z_1,\ldots,Z_{k-1},Z_{k+1},\ldots,Z_{11},
Z_{12},Z_{13})\right. \\ \nn && \left. +4q\sin^2\varphi \
\gamma_{max}(1-\gamma_{max})\
\left(\frac{C_{\phi_{2-}}}{(2\gamma_{max}-1)^{2}(1-\gamma_{max}(1-\sin\varphi))^{2}}\times\right.\right.
\\ \nn && \left.\left.
F_D^{(11)}(1/2;b_2,\ldots,b_{k-1},b_{k+1},\ldots,b_{11},b_{12},b_{13};1;Z_2,\ldots,Z_{k-1},Z_{k+1},\ldots,Z_{11},
Z_{12},Z_{13})\right.\right. \\ \nn &&\left.\left.
+\frac{4q\alpha\Lambda^{\phi_{1-}}\gamma_{max}(1-\gamma_{max})}{(2\gamma_{max}-1)^{2}(1-\gamma_{max}(1-\sin\varphi))^{4}}
\times\right.\right.
\\ \nn &&\left.\left.  F_D^{(12)}(1/2;2,1,b_3,\ldots,b_{k-1},b_{k+1},\ldots,b_{11},-2,-2;1;Z_1,\ldots,Z_{k-1},Z_{k+1},\ldots,Z_{11},
Z_{12},Z_{13})\right)\right], \eea

\bea\label{J2mf} &&J_{2-}=\frac{2\pi
T}{\sin\varphi}(-a_{10})^{-1/2}
(1-\gamma_{max}(1-\sin\varphi))^{2}(2\gamma_{max}-1)(\gamma_{max}-\gamma_{min})^{1/2}
\\ \nn
&&\left[\prod_{i=2}^{10}(\gamma_{max}-\gamma_i)\right]^{-1/2}\times
\\ \nn &&
\left[\frac{\beta}{\alpha}C_{\phi_{2-}}
F_D^{(10)}(1/2;b_1,\ldots,b_{k-1},b_{k+1},\ldots,b_{11};1;Z_1,\ldots,Z_{k-1},Z_{k+1},\ldots,Z_{11})\right.
\\ \nn
&&\left.+\frac{\alpha\Lambda^{\phi_{2-}}}{\sin^2\varphi}(2\gamma_{max}-1)^{2}\times\right.
\\ \nn &&\left.
F_D^{(10)}(1/2;b_1,-3,b_3,\ldots,b_{k-1},b_{k+1},\ldots,b_{11};1;Z_1,\ldots,Z_{k-1},Z_{k+1},\ldots,Z_{11})\right.
\\ \nn &&\left. -q\sin^2\varphi \
C_{\phi_{1-}}(1-\gamma_{max}(1-\sin\varphi))\times\right.
\\ \nn &&\left.
F_D^{(9)}(1/2;b_2,\ldots,b_{k-1},b_{k+1},\ldots,b_{11};1;Z_2,\ldots,Z_{k-1},Z_{k+1},\ldots,Z_{11})\right.
\\ \nn &&\left. +4q^2\alpha\Lambda^{\phi_{2-}}\sin^2\varphi \ \gamma_{max}(1-\gamma_{max})
(1-\gamma_{max}(1-\sin\varphi))^{-4}\times\right. \\ \nn && \left.
F_D^{(13)}(1/2;2,-1,b_3,\ldots,b_{k-1},b_{k+1},\ldots,b_{11},-1,-1;1;Z_1,\ldots,Z_{k-1},Z_{k+1},\ldots,Z_{11},
Z_{12},Z_{13})\right],\eea where $Z_k$ are related to the previous
$z_k$ by the change $\gamma\rightarrow\gamma_{min}$.

In writing (\ref{J1mf}), (\ref{J2mf}), we used the following
property of the hypergeometric functions $F_D^{(n)}$:
\bea\label{PBMv3P}
&&F_D^{(n)}(a;b_1,\ldots,b_n;c;z_1,\ldots,z_{k-1},1,z_{k+1},\ldots,z_n)=
\frac{\Gamma(c)\Gamma(c-a-b_k)}{\Gamma(c-a)\Gamma(c-b_k)}\times
\\ \nn &&
F_D^{(n-1)}(a;b_1,\ldots,b_{k-1},b_{k+1},\ldots,b_n;c-b_k;z_1,\ldots,z_{k-1},z_{k+1},\ldots,z_n).\eea
It follows from the integral representation (\ref{FDn}). We needed
to use this property in order to take into account that for some
$k$ \bea\nn
Z_k=\frac{\gamma_{max}-\gamma_{min}}{\gamma_{max}-\gamma_{k-1}}=1
,\eea i.e. $\gamma_{k-1}=\gamma_{min}\ge 0$.

\setcounter{equation}{0}
%%%%%%%%%%%%%5%%%%%%%%%%%%%%%%%%%%%%%%%%%%%%%%%%%%%%
\section{Concluding remarks}
%%%%%%%%%%%%%5%%%%%%%%%%%%%%%%%%%%%%%%%%%%%%%%%%%%%%

Here we considered strings living in $AdS_3\times S^3\times
S^3\times S^1$ with nonzero 2-form $B$-field. By using specific
ansatz for the string embedding, we obtained a class of solutions
corresponding to strings moving in the whole ten dimensional
space-time. For the $AdS_3$ subspace, these solutions are given in
terms of incomplete elliptic integrals as expected. For the two
three-spheres, they are expressed in terms of Lauricella
hypergeometric functions of many variables. The same is true for
the corresponding conserved angular momenta related to the
isometries of the three-spheres. This is in contrast with the case
of $AdS_3\times S^3\times T^4$ background with $B$-field, where
the solutions for the string coordinates on $S^3$ are given in
terms of incomplete elliptic integrals \cite{AB1404,BPS1508}. The
complications here arise because of the specific form of the
$B$-field for this supergravity solution (see (\ref{Bour})).

\section*{Acknowledgements}
This work is partially supported by the NSF grant DFNI T02/6.

\section*{Appendix}
\def\theequation{A.\arabic{equation}}
\setcounter{equation}{0}
\begin{appendix}
Here we explain how (\ref{dxi1}) is derived. First we represent
$d\xi$ as \bea\nn d\xi =\frac{\alpha^{2}-\beta^{2}}{\sin\varphi}
\left(1-(1-\sin\varphi)\gamma\right)^{2}(2\gamma-1)
\sum_{i=0}^{10}a_i \gamma^{i}d\gamma,\eea where \bea\nn
a_0=-\frac{1}{4}C_{\phi_{1-}}^2\sin^2\varphi, \eea \bea\nn a_1&=&
C_{\theta_-}(\alpha^{2}-\beta^{2})
-\alpha^{2}\left(\Lambda^{\phi_{2-}}\right)^{2}\csc^2\varphi
\\ \nn &&
-\sin^2\varphi \left[\left(C_{\phi_{1-}}\right)^{2}\sin\varphi +
\left(C_{\phi_{2-}}\right)^{2}
-2C_{\phi_{1-}}\left(C_{\phi_{1-}}+q\alpha\Lambda^{\phi_{2-}}\right)\right],\eea
\bea\nn a_2&=&
-9C_{\theta_-}(\alpha^{2}-\beta^{2})-4\alpha^{2}\left(\Lambda^{\phi_{2-}}\right)^{2}\csc\varphi
+\alpha^{2}\left[13\left(\Lambda^{\phi_{2-}}\right)^{2}
-4\left(\Lambda^{\phi_{1-}}\right)^{2}\right]\csc^2\varphi
\\ \nn && +\frac{1}{2}\sin\varphi \left(8C_{\theta_-}(\alpha^{2}-\beta^{2})
-\sin\varphi\left(13\left(C_{\phi_{1-}}\right)^{2}-2C_{\phi_{2-}}\left(5C_{\phi_{2-}}
-8q\alpha\Lambda^{\phi_{1-}}\right)\right.\right.
\\ \nn
&&\left.\left.+28q\alpha
C_{\phi_{1-}}\Lambda^{\phi_{2-}}+8q^2\alpha^2\left(\Lambda^{\phi_{2-}}\right)^{2}
+\sin\varphi\left(8\left(C_{\phi_{2-}}\right)^{2}\right.\right.\right. \\
\nn &&\left.\left.\left.
-2C_{\phi_{1-}}\left(7C_{\phi_{1-}}+4q\alpha
\Lambda^{\phi_{2-}}\right)+3\left(C_{\phi_{1-}}\right)^{2}\sin\varphi\right)
\right)\right),\eea \bea\nn a_3&=&
34C_{\theta_-}(\alpha^{2}-\beta^{2})
-6\alpha^{2}\left(\Lambda^{\phi_{2-}}\right)^{2}-16\alpha^{2}\left(\left(\Lambda^{\phi_{1-}}\right)^{2}
-3\left(\Lambda^{\phi_{2-}}\right)^{2}\right)\csc\varphi \\
\nn &&+2\alpha^{2}\left(20\left(\Lambda^{\phi_{1-}}\right)^{2}
-37\left(\Lambda^{\phi_{2-}}\right)^{2}\right)\csc^2\varphi
+\sin\varphi\times \\ \nn
&&\left(-32C_{\theta_-}(\alpha^{2}-\beta^{2})
+\sin\varphi\left(11\left(C_{\phi_{1-}}\right)^{2} +38q\alpha
C_{\phi_{1-}}\Lambda^{\phi_{2-}}\right.\right. \\ \nn
&&\left.\left.+2\left(3C_{\theta_-}(\alpha^{2}-\beta^{2})
-5\left(C_{\phi_{2-}}\right)^{2}+16q\alpha
C_{\phi_{2-}}\Lambda^{\phi_{1-}}\right.\right.\right.
\\ \nn &&\left.\left.\left. -4q^2\alpha^{2}\left(2\left(\Lambda^{\phi_{1-}}\right)^{2}
-3\left(\Lambda^{\phi_{2-}}\right)^{2}\right)\right) -\sin\varphi
\left(19\left(C_{\phi_{1-}}\right)^{2}\right.\right.\right.
\\ \nn &&\left.\left.\left.-16C_{\phi_{2-}}\left(C_{\phi_{2-}}-q\alpha\Lambda^{\phi_{1-}}\right)
+24q\alpha C_{\phi_{1-}}\Lambda^{\phi_{2-}}
+\sin\varphi\times\right.\right.\right.
\\ \nn &&\left.\left.\left. \left(6\left(C_{\phi_{2-}}\right)^{2}
-9\left(C_{\phi_{1-}}\right)^{2}+\left(C_{\phi_{1-}}\right)^{2}\sin\varphi
-2q\alpha
C_{\phi_{1-}}\Lambda^{\phi_{2-}}\right)\right)\right)\right),\eea
\bea \nn a_4&=& -70C_{\theta_-}(\alpha^{2}-\beta^{2})
-6\alpha^{2}\left(4\left(\Lambda^{\phi_{1-}}\right)^{2}
-11\left(\Lambda^{\phi_{2-}}\right)^{2}\right) \\ \nn &&+
8\alpha^{2}\left(18\left(\Lambda^{\phi_{1-}}\right)^{2}-31\left(\Lambda^{\phi_{2-}}\right)^{2}\right)
\csc\varphi
-2\alpha^{2}\left(86\left(\Lambda^{\phi_{1-}}\right)^{2}-121\left(\Lambda^{\phi_{2-}}\right)^{2}\right)
\csc^2\varphi \\ \nn &&+\frac{1}{4}\sin\varphi
\left(416C_{\theta_-}(\alpha^{2}-\beta^{2})
-16\alpha^{2}\left(\Lambda^{\phi_{2-}}\right)^{2}
+\sin\varphi\times\right. \\ \nn &&\left.
\left(-41\left(C_{\phi_{1-}}\right)^{2}-200q\alpha
C_{\phi_{1-}}\Lambda^{\phi_{2-}}
-8\left(21C_{\theta_-}(\alpha^{2}-\beta^{2})
-5\left(C_{\phi_{2-}}\right)^{2}\right.\right.\right. \\
\nn &&\left.\left.\left.+24q\alpha
C_{\phi_{2-}}\Lambda^{\phi_{1-}}
-2q^2\alpha^2\left(\left(12\Lambda^{\phi_{1-}}\right)^{2}-13\left(\Lambda^{\phi_{2-}}\right)^{2}
\right)\right)+\sin\varphi\times\right.\right.
\\ \nn &&\left.\left.\left(4\left(25\left(C_{\phi_{1-}}\right)^{2}
+4\left(C_{\theta_-}(\alpha^{2}-\beta^{2})-6C_{\phi_{2-}}\left(C_{\phi_{2-}}-2q\alpha
\Lambda^{\phi_{1-}}\right)\right)\right.\right.\right.\right.
\\ \nn &&\left.\left.\left.\left.+52q\alpha C_{\phi_{1-}}\Lambda^{\phi_{2-}}\right)
-\sin\varphi\left(78\left(C_{\phi_{1-}}\right)^{2}
-8C_{\phi_{2-}}\left(9C_{\phi_{2-}}-4q\alpha\Lambda^{\phi_{1-}}\right)\right.\right.\right.\right.
\\ \nn &&\left.\left.\left.\left.+40q\alpha C_{\phi_{1-}}\Lambda^{\phi_{2-}}
+\sin\varphi\left(\left(C_{\phi_{1-}}\right)^{2}\sin\varphi-20\left(C_{\phi_{1-}}\right)^{2}
+16\left(C_{\phi_{2-}}\right)^{2}\right)\right)\right)\right)\right),\eea
\bea \nn a_5&=& 85C_{\theta_-}(\alpha^{2}-\beta^{2})
+6\alpha^{2}\left(32\left(\Lambda^{\phi_{1-}}\right)^{2}
-51\left(\Lambda^{\phi_{2-}}\right)^{2}\right) \\ \nn
&&-16\alpha^2\left(34\left(\Lambda^{\phi_{1-}}\right)^{2}-45\left(\Lambda^{\phi_{2-}}\right)^{2}\right)\csc\varphi
+\alpha^2
\left(416\left(\Lambda^{\phi_{1-}}\right)^{2}-501\left(\Lambda^{\phi_{2-}}\right)^{2}\right)\csc^2\varphi
\\ \nn &&+\sin\varphi\left(-8\left(22C_{\theta_-}(\alpha^{2}-\beta^{2})
+\alpha^2\left(2\left(\Lambda^{\phi_{1-}}\right)^{2}-5\left(\Lambda^{\phi_{2-}}\right)^{2}\right)\right)\right.
\\ \nn &&\left.+\sin\varphi\left(5\left(C_{\phi_{1-}}\right)^{2}+114C_{\theta_-}(\alpha^{2}-\beta^{2})
-\left(5C_{\phi_{2-}}-12q\alpha\Lambda^{\phi_{1-}}\right)
\left(C_{\phi_{2-}}-4q\alpha\Lambda^{\phi_{1-}}\right)\right.\right.
\\ \nn &&\left.\left.+32q\alpha C_{\phi_{1-}}\Lambda^{\phi_{2-}}
-\left(1-48q^2\right)\alpha^2\left(\Lambda^{\phi_{2-}}\right)^{2}
+\sin\varphi\left(-8\left(3C_{\theta_-}(\alpha^{2}-\beta^{2})\right.\right.\right.\right.
\\ \nn &&\left.\left.\left.\left.-2\left(C_{\phi_{2-}}\right)^{2}
+6q\alpha
C_{\phi_{2-}}\Lambda^{\phi_{1-}}+2C_{\phi_{1-}}\left(C_{\phi_{1-}}
+3q\alpha\Lambda^{\phi_{2-}} \right)\right)\right.\right.\right.
\\ \nn &&\left.\left.\left.+\sin\varphi\left(C_{\theta_-}(\alpha^{2}-\beta^{2})
-2C_{\phi_{2-}}\left(9C_{\phi_{2-}}-8q\alpha
\Lambda^{\phi_{1-}}\right)\right.\right.\right.\right.
\\ \nn &&\left.\left.\left.\left. +2C_{\phi_{1-}}
\left(9C_{\phi_{1-}}+8q\alpha\Lambda^{\phi_{2-}}\right)
\right.\right.\right.\right.
\\ \nn &&\left.\left.\left.\left.
+\frac{1}{2}\left(\left(C_{\phi_{1-}}\right)^{2}-\left(C_{\phi_{2-}}\right)^{2}
 \right)\left(1-\cos 2\varphi-16\sin\varphi\right)\right)
\right)\right)\right),\eea \bea \nn a_6&=&-61
C_{\theta_-}(\alpha^{2}-\beta^{2})
-6\alpha^{2}\left(104\left(\Lambda^{\phi_{1-}}\right)^{2}
-129(\left(\Lambda^{\phi_{2-}}\right)^{2}\right) \\ \nn
&&+4\alpha^2 \left(280\left(\Lambda^{\phi_{1-}}\right)^{2}
-321\left(\Lambda^{\phi_{2-}}\right)^{2}\right)\csc\varphi \\ \nn
&&-\alpha^2\left(620\left(\Lambda^{\phi_{1-}}\right)^{2}
-681\left(\Lambda^{\phi_{2-}}\right)^{2}\right)\csc^2\varphi
+\sin\varphi\times
\\ \nn &&\left(4\left(41C_{\theta_-}(\alpha^{2}-\beta^{2})
+\alpha^2\left(28 \left(\Lambda^{\phi_{1-}}\right)^{2}
-41\left(\Lambda^{\phi_{2-}}\right)^{2}\right)\right)
+\sin\varphi\times\right.
\\ \nn &&\left.
\left(-150C_{\theta_-}(\alpha^{2}-\beta^{2})
+\left(C_{\phi_{2-}}+2(1-2q)\alpha\Lambda^{\phi_{1-}}\right)
\left(C_{\phi_{2-}}-2(1+2q)\alpha\Lambda^{\phi_{1-}}\right)\right.\right.
\\ \nn &&\left.\left.-\left(C_{\phi_{1-}}-(3-4q)\alpha\Lambda^{\phi_{2-}}\right)
\left(C_{\phi_{1-}}+(3+4q)\alpha\Lambda^{\phi_{2-}}\right)+\sin\varphi\times\right.\right.
\\ \nn &&\left.\left.\left(4\left(13C_{\theta_-}(\alpha^{2}-\beta^{2})+\left(C_{\phi_{1-}}\right)^{2}-\left(C_{\phi_{2-}}\right)^{2}
+4q\alpha\left(C_{\phi_{1-}}\Lambda^{\phi_{2-}}
+C_{\phi_{2-}}\Lambda^{\phi_{1-}}\right)\right.\right.\right.\right.
\\ \nn &&\left.\left.\left.\left.-\sin\varphi\left(5C_{\theta_-}(\alpha^{2}-\beta^{2})
+6\left(\left(C_{\phi_{1-}}\right)^{2}-\left(C_{\phi_{2-}}\right)^{2}
\right)\right.\right.\right.\right.\right.
\\ \nn &&\left.\left.\left.\left.\left.+8q\alpha\left(C_{\phi_{1-}}\Lambda^{\phi_{2-}}
+C_{\phi_{2-}}\Lambda^{\phi_{1-}}\right)
+\frac{1}{2}\left(\left(C_{\phi_{1-}}\right)^{2}-\left(C_{\phi_{2-}}\right)^{2}
\right)\times\right.\right.\right.\right.\right.
\\ \nn &&\left.\left.\left.\left.\left.\left(1-\cos 2\varphi-8\sin\varphi\right)\right)\right)\right)\right)\right),\eea
\bea \nn a_7&=& -8(1-\sin\varphi)
\left(-3C_{\theta_-}(\alpha^{2}-\beta^{2})
-5\alpha^2\left(7\left(\Lambda^{\phi_{1-}}\right)^{2}
-8\left(\Lambda^{\phi_{2-}}\right)^{2}\right)\right.
\\ \nn &&\left.+97\alpha^{2}\left(97\left(\Lambda^{\phi_{1-}}\right)^{2}
-104\left(\Lambda^{\phi_{2-}}\right)^{2}\right)\csc\varphi
-\alpha^{2}\left(73\left(\Lambda^{\phi_{1-}}\right)^{2}
-76\left(\Lambda^{\phi_{2-}}\right)^{2}\right)
\csc^2\varphi\right.
\\ \nn &&\left.
+\sin\varphi\left(7C_{\theta_-}(\alpha^{2}-\beta^{2})+\alpha^{2}
\left(3\left(\Lambda^{\phi_{1-}}\right)^{2}
-4\left(\Lambda^{\phi_{2-}}\right)^{2}\right)\right.\right.
\\ \nn &&\left.\left.+\frac{1}{2}C_{\theta_-}(\alpha^{2}-\beta^{2}) \left(1-\cos
2\varphi -10\sin\varphi\right)\right)\right) ,\eea \bea \nn a_8&=&
-4\left(1-\sin\varphi\right)^{2}
\left(C_{\theta_-}(\alpha^{2}-\beta^{2})+\alpha^2\left(13\left(\Lambda^{\phi_{1-}}\right)^{2}
-14\left(\Lambda^{\phi_{2-}}\right)^{2}\right)\right.
\\ \nn &&\left.-2\alpha^2\left(37\left(\Lambda^{\phi_{1-}}\right)^{2}
-38\left(\Lambda^{\phi_{2-}}\right)^{2}\right)\csc\varphi\right.
\\ \nn &&\left.+\alpha^2\left(85\left(\Lambda^{\phi_{1-}}\right)^{2}
-86\left(\Lambda^{\phi_{2-}}\right)^{2}\right)\csc^2\varphi\right.
\\ \nn &&\left.+\frac{1}{2}C_{\theta_-}(\alpha^{2}-\beta^{2}) \left(1-\cos
2\varphi -4\sin\varphi\right)\right),\eea \bea \nn a_9&=&
16\alpha^2\left(\left(\Lambda^{\phi_{1-}}\right)^{2}
-\left(\Lambda^{\phi_{2-}}\right)^{2}\right)\csc^2\varphi
\left(1-\sin\varphi\right)^{3} \left(7-3\sin\varphi\right),\eea
\bea\nn a_{10}&=&
-16\alpha^2\left(\left(\Lambda^{\phi_{1-}}\right)^{2}
-\left(\Lambda^{\phi_{2-}}\right)^{2}\right)\csc^2\varphi
\left(1-\sin\varphi\right)^{4}.\eea

Since we want the variable $\gamma=\cos^2\frac{\theta_-}{2}$ to
have {\it maximum}, the coefficient $a_{10}$ must be negative,
i.e. \bea\nn \left(\Lambda^{\phi_{1-}}\right)^{2}
>\left(\Lambda^{\phi_{2-}}\right)^{2}.\eea
Taking this into account, we rewrite $\sum_{i=0}^{10}a_i
\gamma^{i}$ as \bea\nn \sum_{i=0}^{10}a_i
\gamma^{i}&=&-a_{10}\left(-\gamma^{10}
-\frac{1}{a_{10}}\sum_{j=0}^{9}a_j \gamma^{j} \right) \equiv
-a_{10}\left(-\gamma^{10} -\sum_{j=0}^{9}B_j \gamma^{j} \right)
\\\nn &=&-a_{10}
(\gamma_1-\gamma)\prod_{k=2}^{10}(\gamma-\gamma_k),\eea where
$\gamma_1=\gamma_{max}$ and \bea\nn -B_0=-\prod_{k=1}^{10}\gamma_k
\h ,\ldots ,\h -B_9= \sum_{k=1}^{10}\gamma_k.\eea

\end{appendix}

\end{document}